\documentclass[preprintnumbers,amsmath,amssymb,showpacs]{revtex4}
\usepackage{graphicx}
\usepackage{dcolumn}
\usepackage{bm}
\usepackage{booktabs}
\usepackage{CJK}
\usepackage{subfigure}
\usepackage{qcircuit}
\usepackage{color}
\usepackage{braket}
\usepackage{listings}
\usepackage{soul}
\usepackage{xcolor}
\usepackage{threeparttable} 
\usepackage{booktabs} 
\usepackage{multirow} 
\soulregister\cite7 
\soulregister\ref7 
\usepackage{color, soul} 
\setstcolor{yellow}

\begin{document}

\title{A more generalized two-qubit symmetric quantum joint measurement}

\author{Ying-Qiu He$^{1}$}
\author{Dong Ding$^{1}$}
\email{dingdong@ncist.edu.cn}
\author{Ting Gao$^{2}$}
\email{gaoting@hebtu.edu.cn}
\author{Zan-Jia Li$^{3}$}
\author{Feng-Li Yan$^{3}$}
\email{flyan@hebtu.edu.cn}

\affiliation {
$^1$ College of Science, North China Institute of Science and Technology, Beijing 101601, China\\
$^2$ School of Mathematical Sciences, Hebei Normal University, Shijiazhuang 050024, China\\
$^3$ College of Physics, Hebei Normal University, Shijiazhuang 050024, China
}
\date{\today}

\begin{abstract}
A standard two-qubit joint measurement is the well-known Bell state measurement (BSM), in which each reduced state (traced out one qubit) is the completely mixed state. Recently, a novel quantum joint measurement named elegant joint measurement (EJM) has been proposed, where the reduced states of the EJM basis have tetrahedral symmetry. In this work, we first suggest a five-parameter entangled state and reveal its inherent symmetry. Based on this, we define a more generalized EJM parameterized by $z$, $\varphi_{}$ and $\theta_{}$, and provide the quantum circuits for preparing and detecting these basis states. There are three main results: (i) the previous single-parameter EJM can be directly obtained by specifying the parameters $z$ and $\varphi_{}$; (ii) the initial unit vectors related to the four vertices of the regular tetrahedron are not limited to the original choice and not all the unit vectors in cylindrical coordinates are suitable for forming the EJM basis; and (iii) the reduced states of the present EJM basis can always form two mirrorimage tetrahedrons, robustly preserving its elegant properties. We focus on figuring out what kind of states the EJM basis belongs to and providing a method for constructing the more generalized three-parameter EJM, which may contribute to the multi-setting measurement and the potential applications for quantum information processing.
\end{abstract}



%
\maketitle

\section{Introduction}

As a resource, quantum entanglement \cite{Quantum-entanglement,Entanglement-detection2009} has two basic roles in the field of quantum information processing \cite{Gisin2014,QST2017}, providing quantum channel and contributing quantum joint measurement.
A typical quantum joint measurement is the well-known Bell state measurement (BSM), consisting of four orthogonal two-qubit entangled states. In a sense, it laid the foundations for quantum computing, and consequently for the quantum information science \cite{NC2000,IQIS2015}.

In 2019, Gisin \cite{Gisin-EJM2019} proposed a novel two-qubit joint measurement inspired by the four vertices of the tetrahedron inscribed in the Poincar\'{e} sphere, named elegant joint measurement (EJM). It has elegant symmetry property since the reduced states of the EJM basis, described by tracing out one party, can also form two mirrorimage regular tetrahedrons of radius $\sqrt{3}/2$ inside the Bloch sphere. More recently, Tavakoli \emph{et al.} \cite{TGB-EJM2021} extended Gisin's original EJM to the generalized EJM by considering a real parameter $\theta \in [0,\pi/2]$, where the elegant symmetry holds and the tetrahedrons are shrunk by $(\sqrt{3}/2)\cos\theta$.
An important application of the EJM is to reveal noise-tolerant nonbilocality in networks \cite{Tavakoli-network2022}, and it has been experimentally realized by superconducting quantum processors \cite{BGT2021IBM} or hyperentanglement photons \cite{GuoPRL-EJM2022}.
In previous works, all the unit vectors, related to the four vertices of the regular tetrahedron, were precisely specified as $\vec{m}_{0}= (1,1,1)/\sqrt{3}, \vec{m}_{1}= (1,-1,-1)/\sqrt{3}, \vec{m}_{2}= (-1,1,-1)/\sqrt{3}$ and $\vec{m}_{3}= (-1,-1,1)/\sqrt{3}$.
In fact, however, an arbitrary unit vector in cylindrical coordinates generally contains two parameters. With adding one more real phase $\theta$, the generalized EJM basis may contain nine parameters.
So, a basic problem is how to describe the generalized EJM basis as few parameters as possible, for simplicity and without loss of generality?

In this work, we focus on the more generalized EJM and consider the following three tasks.
Firstly, we define a five-parameter two-qubit entangled state based on an arbitrary unit vector in cylindrical coordinates. We then calculate the concurrence and the partial traces to reveal its inherent symmetry.
Secondly, we define the more generalized EJM via the present five-parameter entangled state.
We want to determine whether these vectors in previous works are uniquely specified and whether all unit vectors in cylindrical coordinates are suitable for constructing the generalized EJM.
Thirdly, we provide the quantum circuits for preparing and detecting the present generalized EJM basis states.

\section{The two-qubit entangled state with elegant symmetry}

\subsection{Definition of the state}

In cylindrical coordinates, one can define an arbitrary unit vector
\begin{eqnarray} \label{vector-m}
 \vec{m} =(\sqrt{1-z^{2}}\cos \varphi, \sqrt{1-z^{2}}\sin \varphi, z),
\end{eqnarray}
related to the qubit state \cite{Gisin-EJM2019,TGB-EJM2021}
\begin{eqnarray}
|m^{}_{}\rangle=\frac{1}{\sqrt{2}}
(\sqrt{1 + z}\text{e}^{-\text{i}\varphi_{}/2}|0\rangle + \sqrt{1 - z}\text{e}^{\text{i}\varphi_{}/2}|1\rangle)
\end{eqnarray}
in the Bloch sphere picture, where the real coefficient $|z| \leq 1$ and phase factor $\varphi \in [-\pi, \pi]$.
Then, let
\begin{eqnarray}
|-m^{}_{}\rangle=\frac{1}{\sqrt{2}}
(\sqrt{1 - z}\text{e}^{-\text{i}\varphi_{}/2}|0\rangle - \sqrt{1 + z}\text{e}^{\text{i}\varphi_{}/2}|1\rangle),
\end{eqnarray}
obviously the states $|m^{}_{}\rangle$ and $|- m^{}_{}\rangle$ are orthogonal.

Here, we further define two states
\begin{eqnarray}
|m^{}_{0}\rangle=\frac{1}{2}[(1-\text{i}\text{e}^{\text{i}\theta_{0}})|m_{}\rangle
                             +(1+\text{i}\text{e}^{\text{i}\theta_{0}})|-m_{}\rangle]
\end{eqnarray}
and
\begin{eqnarray}
|m^{}_{1}\rangle=\frac{1}{2}[(1+\text{i}\text{e}^{\text{i}\theta_{0}})|m_{}\rangle
                             +(1-\text{i}\text{e}^{\text{i}\theta_{0}})|-m_{}\rangle]
\end{eqnarray}
by introducing a real parameter $\theta_{0} \in [0,\pi/2]$.
It is easy to see that these two states are orthogonal, i.e., $\langle m^{}_{i}|m^{}_{j}\rangle=\delta_{ij}, i,j=0,1$, where $\delta$ is the Kronecker symbol.
Note that for $\theta_{0} = \pi/2$ we have $|m^{}_{0}\rangle=|m_{}\rangle$ and $|m^{}_{1}\rangle=|-m_{}\rangle$.

Now, in the basis $\{|m^{}_{0}\rangle, |m^{}_{1}\rangle\}$, we define a two-qubit state
\begin{eqnarray} \label{EJM-state}
|\Phi_{}\rangle=\frac{1}{\sqrt{2a^{2}+2}}  [(a+\text{e}^{\text{i}\theta_{}})|m_{0},m_{1}\rangle
                                     +(a-\text{e}^{\text{i}\theta_{}})|m_{1},m_{0}\rangle],
\end{eqnarray}
where $a$ is an arbitrary real parameter and $\theta_{} \in [0,\pi/2]$,
\begin{eqnarray}
|m^{}_{0},m^{}_{1}\rangle=\frac{1}{4}[(1+\text{e}^{2\text{i}\theta_{0}})|m_{},m_{}\rangle
                                     +(1-\text{i}\text{e}^{\text{i}\theta_{0}})^{2}|m_{},-m_{}\rangle
                                     +(1+\text{i}\text{e}^{\text{i}\theta_{0}})^{2}|-m_{},m_{}\rangle
                                     +(1+\text{e}^{2\text{i}\theta_{0}})|-m_{},-m_{}\rangle]
\end{eqnarray}
and
\begin{eqnarray}
|m^{}_{1},m^{}_{0}\rangle=\frac{1}{4}[(1+\text{e}^{2\text{i}\theta_{0}})|m_{},m_{}\rangle
                                     +(1+\text{i}\text{e}^{\text{i}\theta_{0}})^{2}|m_{},-m_{}\rangle
                                     +(1-\text{i}\text{e}^{\text{i}\theta_{0}})^{2}|-m_{},m_{}\rangle
                                     +(1+\text{e}^{2\text{i}\theta_{0}})|-m_{},-m_{}\rangle].
\end{eqnarray}
It contains five real parameters in the state (\ref{EJM-state}), i.e., $a,z,\varphi$, $\theta_{0}$ and $\theta_{}$.
We will demonstrate two main properties of the present state and then construct a set of generalized two-qubit joint measurement basis involving three continuous parameters $z$, $\varphi$ and $\theta_{}$.

\subsection{The elegant properties}

One can rewrite the present state (\ref{EJM-state}) by using the computational basis, as
\begin{eqnarray}
|\Phi_{}\rangle &=& \frac{1}{2\sqrt{a^{2}+1}}  [a(r_{+}+\sqrt{1-z^{2}}r_{-})\text{e}^{-\text{i}\varphi}|00\rangle
                                          - (azr_{-}-\sqrt{2}\text{i}\text{e}^{\text{i}(\theta_{0}+\theta_{})}) |01\rangle
                                          - (azr_{-}+\sqrt{2}\text{i}\text{e}^{\text{i}(\theta_{0}+\theta_{})}) |10\rangle  \nonumber \\
                                        &&   +  a(r_{+}-\sqrt{1-z^{2}}r_{-})\text{e}^{\text{i}\varphi}|11\rangle],
\end{eqnarray}
where
$r_{\pm }^{} =(1\pm \text{e}^{2\text{i}\theta_{0}}) /\sqrt{2}$.

\begin{figure}[h]
\centering
      \subfigure[ ~The concurrence of the state (\ref{EJM-state}) with varied $a$ and $\theta$]
           {
           \includegraphics[width=9cm, height=7cm]{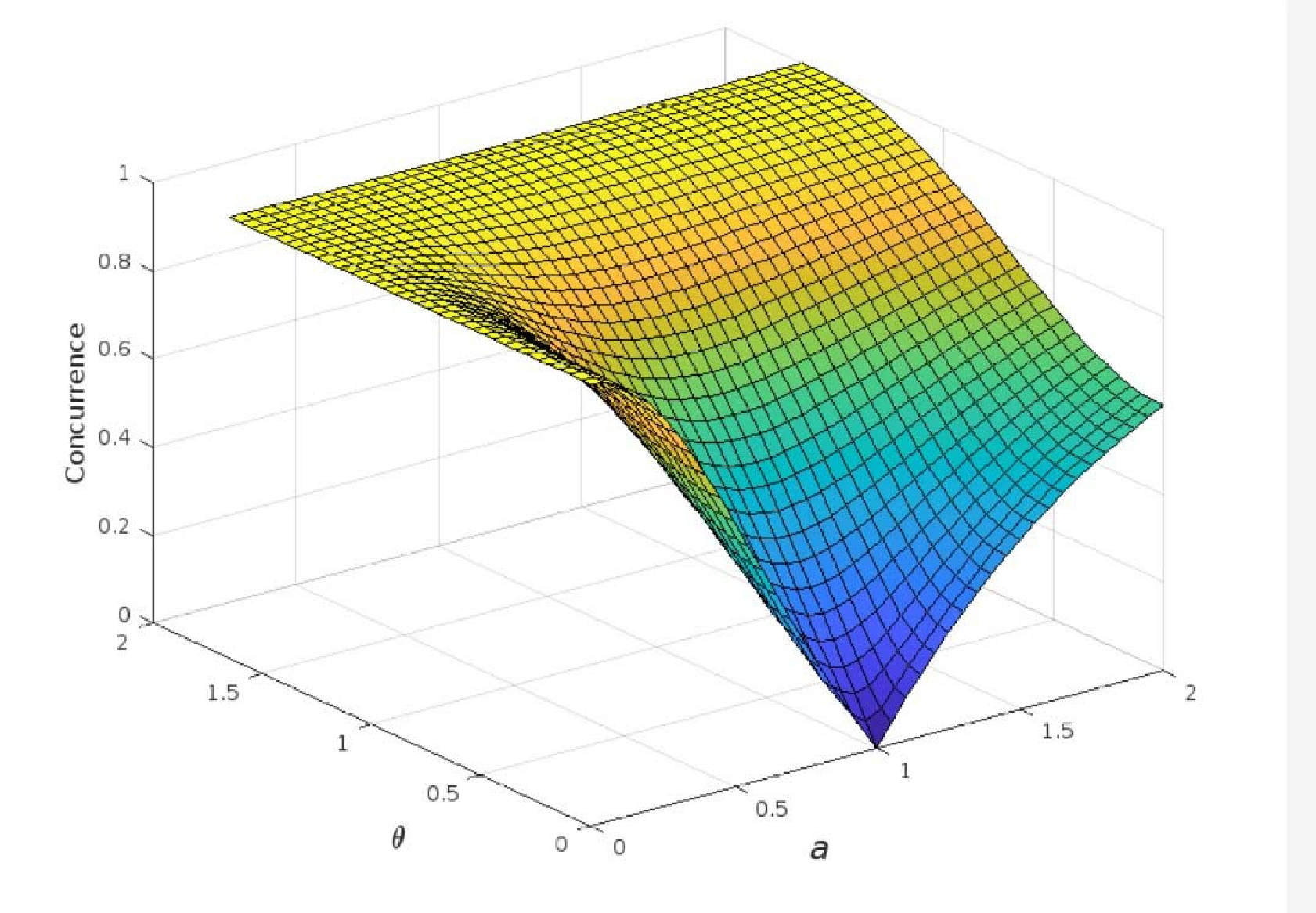}
           }
     \hspace{0.0005in}
      \subfigure[ ~The concurrence of the state (\ref{EJM-state}) with varied $\theta$ for $a=\sqrt{3}$]
           {
           \includegraphics[width=8cm, height=6.5cm]{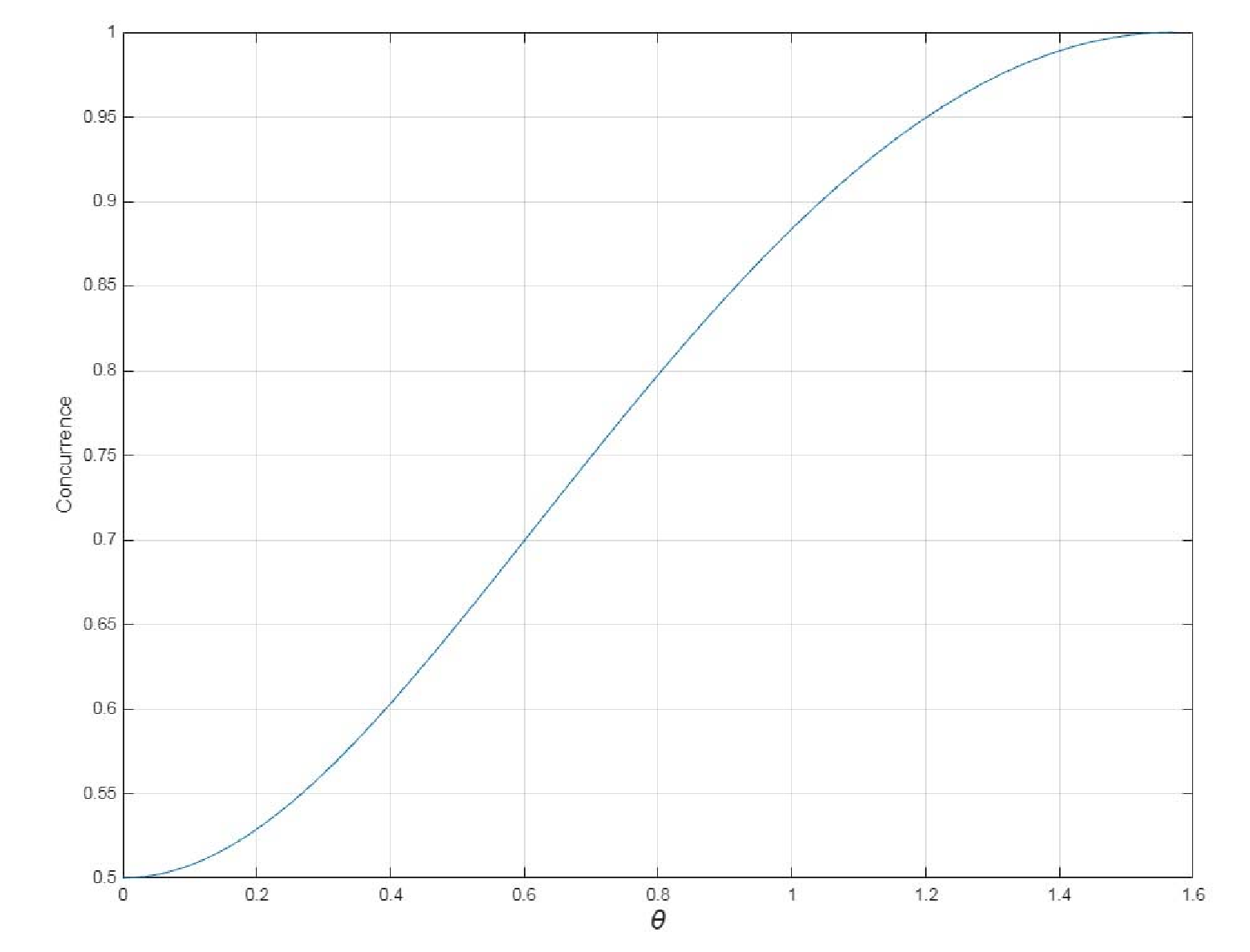}
           }
           \hspace{0.0005in}
  \caption{Plots of the concurrence of the state (\ref{EJM-state}) varied as (a) the parameters $a$ and $\theta$, and (b) the parameter  $\theta$ for a given $a=\sqrt{3}$.}
  \label{concurrence.eps}
\end{figure}

We calculate the concurrence \cite{Quantum-entanglement} by
$C = \sqrt{2(1-\text{tr}\rho_{}^{2})}$, where $\rho_{}$ is a reduced state obtained by tracing out one qubit.
A straightforward calculation shows that
\begin{eqnarray}
C(|\Phi_{}\rangle)  = \sqrt{1-\frac{2 a^{2}(1+\cos 2\theta)}{(a^{2} +1)^{2}}}.
\end{eqnarray}
So the first property of the state (\ref{EJM-state}) is that the concurrence can only depend on the parameters $a$ and $\theta_{}$, rather than all of these parameters.
Note that for $\theta_{} = \pi/2$ or $a=0$ we have $C(|\Phi_{}\rangle)=1$ related to the maximally entangled states (e.g., the singlet state when $a=0$); while for $\theta_{} = 0$ and $a=\pm 1$ one gets the product states $|m_{0},m_{1}\rangle$ or $|m_{1},m_{0}\rangle$ with $C(|\Phi_{}\rangle)  = 0$.
For ease of intuition we plot the concurrence as a function of $a \in [0,2]$ and $\theta \in [0,\pi/2]$, as shown in Fig.\ref{concurrence.eps} (a).
It is straightforward to show that the concurrence varies continuously as $a$ and $\theta$, and the maximum value 1 occurs at $\theta = \pi/2$ or $a=0$, and the minimum zero occurs at $\theta_{} = 0$ and $a=1$.
Also, we provide a plot of the concurrence of the present state for $a=\sqrt{3}$, varied only as $\theta$, as shown in Fig.\ref{concurrence.eps} (b), where $C(|\Phi_{}\rangle) \in [0.5, 1]$.

We next see two reduced states obtained by tracing out one qubit, respectively.

Take $\vec{\sigma}=(\sigma_{x},\sigma_{y},\sigma_{z})$, where $\sigma_{x},\sigma_{y},\sigma_{z}$ are three Pauli observables, and then we calculate $\langle \Phi_{}|\vec{\sigma} \otimes I |\Phi_{}\rangle$ and $\langle \Phi_{}|I \otimes \vec{\sigma}|\Phi_{}\rangle$ to observe its inherent symmetry \cite{Gisin-EJM2019}. We first calculate $\langle \Phi_{}|\sigma_{x} \otimes I |\Phi_{}\rangle$, and have
\begin{eqnarray}
 \langle \Phi_{}|\sigma_{x} \otimes I |\Phi_{}\rangle &=& -\frac{a}{4(a^{2}+1)}[
 (r_{+}^{\ast}+\sqrt{1-z^{2}_{}}r_{-}^{\ast})(az_{}r_{-}+\sqrt{2}\text{i}\text{e}^{\text{i}(\theta_{0}+\theta_{})})\text{e}^{\text{i}\varphi}
 \nonumber \\    &&
+(r_{+}^{\ast}-\sqrt{1-z^{2}_{}}r_{-}^{\ast})(az_{}r_{-}-\sqrt{2}\text{i}\text{e}^{\text{i}(\theta_{0}+\theta_{})})\text{e}^{-\text{i}\varphi}]
+\text{H.c.},
\nonumber \\
 &=& \frac{2a\cos \theta_{}}{a^{2}+1} (\sqrt{1-z^{2}}\cos \varphi \sin\theta_{0}+\sin \varphi \cos \theta_{0}).
\end{eqnarray}
A similar calculation for the other two Pauli observables yields
\begin{eqnarray}
 \langle \Phi_{}|\sigma_{y} \otimes I |\Phi_{}\rangle &=& \frac{2a\cos \theta_{}}{a^{2}+1}  (\sqrt{1-z^{2}}\sin \varphi \sin\theta_{0}-\cos \varphi \cos \theta_{0})
\end{eqnarray}
and
\begin{eqnarray}
 \langle \Phi_{}|\sigma_{z} \otimes I |\Phi_{}\rangle &=& \frac{2a\cos \theta_{}}{a^{2}+1} z \sin\theta_{0}.
\end{eqnarray}
Thus we have
\begin{eqnarray} \label{reduced-state}
 \langle \Phi_{}|\vec{\sigma} \otimes I |\Phi_{}\rangle &=&
\frac{2a\cos \theta_{}}{a^{2}+1} (\sqrt{1-z^{2}}\cos \varphi \sin\theta_{0}+\sin \varphi \cos \theta_{0}, \sqrt{1-z^{2}}\sin \varphi \sin\theta_{0}-\cos \varphi \cos \theta_{0},z\sin \theta_{0}).
\end{eqnarray}
Meanwhile, we calculate $\langle \Phi_{}|I \otimes \vec{\sigma}|\Phi_{}\rangle$, and then obtain the expected elegant property
\begin{eqnarray}
 \langle \Phi_{}|I \otimes \vec{\sigma}|\Phi_{}\rangle &=& -\langle \Phi_{}|\vec{\sigma} \otimes I |\Phi_{}\rangle.
\end{eqnarray}
This is another remarkable property of the present state (\ref{EJM-state}).

Furthermore, one can find that each reduced state is shrunk by $2a\cos \theta /({a^{2}+1})$ inside the Bloch sphere, and towards the point
\begin{eqnarray} \label{vector-m'}
 \vec{m'} =(\sqrt{1-z^{2}}\cos \varphi \sin\theta_{0}+\sin \varphi \cos \theta_{0}, \sqrt{1-z^{2}}\sin \varphi \sin\theta_{0}-\cos \varphi \cos \theta_{0},z\sin \theta_{0})
\end{eqnarray}
or the opposite $- \vec{m'}$.
It matches the largest modulus $2a/({a^{2}+1})$ for $\theta=0$ and has the smallest modulus zero for $\theta=\pi/2$ or $a=0$ related to the completely mixed state.
Specially, if one takes $a=\sqrt{3}$ and $\theta_{0}=\pi/2$, then it is easily checked that $\langle \Phi_{}|\vec{\sigma} \otimes I |\Phi_{}\rangle=(\sqrt{3}/2)\cos \theta_{} (\sqrt{1-z^{2}}\cos \varphi, \sqrt{1-z^{2}}\sin \varphi,z)=(\sqrt{3}/2)\cos \theta_{} \vec{m}$; this corresponds to a three-parameter two-qubit entangled state
\begin{eqnarray} \label{EJM2021}
|\Phi_{}(\theta)\rangle=\frac{1}{2\sqrt{2}}  [(\sqrt{3}+\text{e}^{\text{i}\theta_{}})|m_{},-m_{}\rangle
                                     +(\sqrt{3}-\text{e}^{\text{i}\theta_{}})|-m_{},m_{}\rangle].
\end{eqnarray}
More explicitly, if one further takes $z=1/\sqrt{3}, -1/\sqrt{3}, 1/\sqrt{3}, -1/\sqrt{3}$ and sequentially $\varphi_{}= \pi/4, 3\pi/4, -3\pi/4, -\pi/4$, then these reduced states are respectively pointing to the unit vectors $(\pm1,\pm1,\pm1)/\sqrt{3}$ shrunk by $(\sqrt{3}/2)\cos \theta$, related to two mirrorimage tetrahedrons inside the Bloch sphere \cite{Gisin-EJM2019, TGB-EJM2021}.
Of course, the choice of parameters is not uniquely determined for obtaining such reduced states. For $\theta_{0} = \arcsin (\sqrt{2/3})$, for example, one can take $z_{}=1/\sqrt{2}, -1/\sqrt{2}, 1/\sqrt{2}, -1/\sqrt{2}$ and $\varphi_{}= \pi/2, \pi, -\pi/2, 0$, sequentially, then these reduced states can also point to the vectors $(\pm1,\pm1,\pm1)/\sqrt{3}$ shrunk by $2a\cos \theta /({a^{2}+1})$.
This may lead to the possibility of characterizing the generalized two-qubit joint measurement including more parameters.

\section{The generalized three-parameter quantum joint measurement}

\subsection{The joint measurement basis}

In this section, we focus on the generalized three-parameter quantum joint measurement.

\textit{Theorem}:
Consider three real parameters $z$, $\varphi_{}$ and $\theta_{}$ , where $1/\sqrt{3} \leq |z|\leq 1$, $\varphi \in [-\pi, \pi]$ and $\theta_{} \in [0, \pi/2]$.
Let $\varphi_{0}=\varphi_{}$, $\varphi_{1}=\varphi_{}+\pi/2$, $\varphi_{2}=\varphi_{}-\pi$, $\varphi_{3}=\varphi_{}-\pi/2$, $z_{0}=z_{2}=z$, $z_{1}=z_{3}=-z$.
Define
\begin{eqnarray} \label{EJM-basis}
|\Phi_{i}\rangle=\frac{1}{2\sqrt{2}}  [(\sqrt{3}+\text{e}^{\text{i}\theta_{}})|m_{i,0},m_{i,1}\rangle
+(\sqrt{3}-\text{e}^{\text{i}\theta_{}})|m_{i,1},m_{i,0}\rangle], ~~i=0,1,2,3,
\end{eqnarray}
where
\begin{eqnarray} \label{m-i-0}
|m^{}_{i,0}\rangle=\frac{1}{2}[(1-\text{i}\text{e}^{\text{i}\theta_{0}})|m_{i}\rangle
                             +(1+\text{i}\text{e}^{\text{i}\theta_{0}})|-m_{i}\rangle],
\end{eqnarray}
\begin{eqnarray} \label{m-i-1}
|m^{}_{i,1}\rangle=\frac{1}{2}[(1+\text{i}\text{e}^{\text{i}\theta_{0}})|m_{i}\rangle
                             +(1-\text{i}\text{e}^{\text{i}\theta_{0}})|-m_{i}\rangle],
\end{eqnarray}
$\theta_{0} = \arcsin (1/\sqrt{3z^{2}})$ and
\begin{eqnarray}
|\pm m^{}_{i}\rangle=\frac{1}{\sqrt{2}}   (\sqrt{1 \pm z_{i}}\text{e}^{-\text{i}\varphi_{i}/2}|0\rangle
                                       \pm \sqrt{1 \mp z_{i}}\text{e}^{\text{i}\varphi_{i}/2}|1\rangle).
\end{eqnarray}
Then there exist a set of two-qubit orthonormal basis $\{|\Phi_{i}\rangle\}$, described by the real parameters $z$, $\varphi_{}$ and $\theta_{}$.

\textit{Proof}: We first see the orthogonality.

Substituting (\ref{m-i-0}) and (\ref{m-i-1}) into (\ref{EJM-basis}) yields
\begin{eqnarray}\label{orthonormal-basis}
|\Phi_{i}\rangle &=& \frac{1}{4}  [\sqrt{3}r_{+}|m_{i},m_{i}\rangle
                           + (\sqrt{3}r_{-}-\sqrt{2}\text{i}\text{e}^{\text{i}(\theta_{0}+\theta_{})}) |m_{i},-m_{i}\rangle \nonumber \\ &&
                           + (\sqrt{3}r_{-}+\sqrt{2}\text{i}\text{e}^{\text{i}(\theta_{0}+\theta_{})}) |-m_{i},m_{i}\rangle
                           +  \sqrt{3}r_{+}|-m_{i},-m_{i}\rangle],
\end{eqnarray}
where $r_{\pm }^{} =(1\pm \text{e}^{2\text{i}\theta_{0}}) /\sqrt{2}$.
We can further write these basis states in the computational basis, as
\begin{eqnarray} \label{orthonormal-basis-cb}
|\Phi_{i}\rangle &=& \frac{1}{4}  [\sqrt{3}(r_{+}+\sqrt{1-z^{2}_{i}}r_{-})\text{e}^{-\text{i}\varphi_{i}}|00\rangle
                                          - (\sqrt{3}z_{i}r_{-}-\sqrt{2}\text{i}\text{e}^{\text{i}(\theta_{0}+\theta_{})}) |01\rangle
                                          - (\sqrt{3}z_{i}r_{-}+\sqrt{2}\text{i}\text{e}^{\text{i}(\theta_{0}+\theta_{})}) |10\rangle  \nonumber \\
                                        &&   +  \sqrt{3}(r_{+}-\sqrt{1-z^{2}_{i}}r_{-})\text{e}^{\text{i}\varphi_{i}}|11\rangle].
\end{eqnarray}
So, for any two state vectors $|\Phi_{i}\rangle$ and $|\Phi_{j}\rangle$, one can calculate the inner product
\begin{eqnarray}
\langle \Phi_{i}|\Phi_{j}\rangle &=& \frac{1}{16} [
 3(r_{+}^{\ast}+\sqrt{1-z^{2}_{i}}r_{-}^{\ast})(r_{+}+\sqrt{1-z^{2}_{j}}r_{-})\text{e}^{\text{i}(\varphi_{i}-\varphi_{j})}
+(\sqrt{3}z_{i}r_{-}^{\ast}+\sqrt{2}\text{i}\text{e}^{-\text{i}(\theta_{0}+\theta_{})})(\sqrt{3}z_{j}r_{-}-\sqrt{2}\text{i}\text{e}^{\text{i}(\theta_{0}+\theta_{})})
 \nonumber \\    &&
+(\sqrt{3}z_{i}r_{-}^{\ast}-\sqrt{2}\text{i}\text{e}^{-\text{i}(\theta_{0}+\theta_{})})(\sqrt{3}z_{j}r_{-}+\sqrt{2}\text{i}\text{e}^{\text{i}(\theta_{0}+\theta_{})})
+3(r_{+}^{\ast}-\sqrt{1-z^{2}_{i}}r_{-}^{\ast})(r_{+}-\sqrt{1-z^{2}_{j}}r_{-})\text{e}^{-\text{i}(\varphi_{i}-\varphi_{j})}]
\nonumber \\  &=& \frac{1}{16} \{
 3[|r_{+}|^{2}+(1-z^{2})|r_{-}|^{2}][\text{e}^{\text{i}(\varphi_{i}-\varphi_{j})}+\text{e}^{-\text{i}(\varphi_{i}-\varphi_{j})}]
+ 6z_{i}z_{j}|r_{-}|^{2} +4  \nonumber \\    &&
+3(\sqrt{1-z^{2}_{i}}r_{+}r_{-}^{\ast}+\sqrt{1-z^{2}_{j}}r_{+}^{\ast}r_{-})[\text{e}^{\text{i}(\varphi_{i}-\varphi_{j})}-\text{e}^{-\text{i}(\varphi_{i}-\varphi_{j})}]\}
\nonumber \\  &=& \frac{1}{4} [2\cos (\varphi_{i}-\varphi_{j})+\text{sng}(z_{i}z_{j})+1].
\end{eqnarray}
Here we use $|r_{\pm}|^{2}=1\pm \cos (2\theta_{0})$,
$r_{+}r_{-}^{\ast}+r_{-}r_{+}^{\ast}=0$, and $z^{2}|r_{-}|^{2} = [1-\cos (2\theta_{0})]/(3\sin^{2}\theta_{0}) = 2/3$.

Note that for $\varphi_{i}-\varphi_{j}=\pi$, we have $z_{i}=z_{j}$ and then $\text{sng}(z_{i}z_{j})=1$; alternatively, for $\varphi_{i}-\varphi_{j}=\pi/2$, we have $\text{sng}(z_{i}z_{j})=-1$.
So we can show explicitly that the states (\ref{EJM-basis}) satisfy the orthogonality relation
\begin{eqnarray}
\langle \Phi_{i}|\Phi_{j}\rangle &=& \delta_{ij}, ~~i,j=0,1,2,3.
\end{eqnarray}

Now let us show the completeness relation, i.e., $\sum_{i}|\Phi_{i}\rangle \langle \Phi_{i}| = I, i=0,1,2,3$.

Since $\theta_{0} = \arcsin (1/\sqrt{3z^{2}})$, we have
$\text{e}^{\text{i}\theta_{0}}=(\sqrt{3z^{2}-1}+\text{i})/\sqrt{3z^{2}}$,
$r_{+}=\sqrt{2}-r_{-}$ and $r_{-}=\sqrt{2}(1-\text{i}\sqrt{3z^{2}-1})/(3z^{2})$.
This leads to the further simplification for the basis states (\ref{EJM-basis}), as
\begin{eqnarray} \label{simplified-basis-cb}
|\Phi_{i}\rangle &=& \frac{1-\text{i}\sqrt{3z^{2}-1}}{2\sqrt{3}z^{2}}  (a_{+}\text{e}^{-\text{i}\varphi_{i}}|00\rangle - b_{i,+}^{\theta}|01\rangle -b_{i,-}^{\theta}|10\rangle + a_{-}\text{e}^{\text{i}\varphi_{i}}|11\rangle),
\end{eqnarray}
where $a_{\pm}=(\text{i}\sqrt{3z^{2}-1} \pm \sqrt{1-z^{2}})/\sqrt{2}$ and $b_{i,\pm}^{\theta}=(z_{i} \pm |z|\text{e}^{\text{i}\theta})/\sqrt{2}$.

Define the outer product operators $A^{i}=|\Phi_{i}\rangle \langle \Phi_{i}|, i=0,1,2,3$.
So in matrix representation we have
\begin{eqnarray}\label{}
  A^{i}=\frac{1}{4z^{2}}\left(
  \begin{array}{cccc}
  |a_{+}|^{2}& ~-a_{+}\text{e}^{-\text{i}\varphi_{i}}(b_{i,+}^{\theta})^{*}& ~-a_{+}\text{e}^{-\text{i}\varphi_{i}}(b_{i,-}^{\theta})^{*}& ~a_{+}a_{-}^{*}\text{e}^{-2\text{i}\varphi_{i}}\\
  -a_{+}^{*}\text{e}^{\text{i}\varphi_{i}}b_{i,+}^{\theta}& ~|b_{i,+}^{\theta}|^{2}& ~b_{i,+}^{\theta}(b_{i,-}^{\theta})^{*}& ~-a_{-}^{*}\text{e}^{-\text{i}\varphi_{i}}b_{i,+}^{\theta}\\
  -a_{+}^{*}\text{e}^{\text{i}\varphi_{i}}b_{i,-}^{\theta}& ~b_{i,-}^{\theta}(b_{i,+}^{\theta})^{*}& ~|b_{i,-}^{\theta}|^{2}& ~-a_{-}^{*}\text{e}^{-\text{i}\varphi_{i}}b_{i,-}^{\theta}\\
  a_{+}^{*}a_{-}\text{e}^{2\text{i}\varphi_{i}}& ~-a_{-}\text{e}^{\text{i}\varphi_{i}}(b_{i,+}^{\theta})^{*}& ~-a_{-}\text{e}^{\text{i}\varphi_{i}}(b_{i,-}^{\theta})^{*}& ~|a_{-}|^{2}\\
  \end{array}
  \right),
\end{eqnarray}
with matrix elements $A^{i}_{kl}, k,l=0,1,2,3$.
For the diagonal elements,
since $|a_{\pm}|^{2}=z^{2}$, $|b_{i,\pm}^{\theta}|^{2} = z^{2} \pm z_{i}|z|\cos \theta$ and $\sum_{i}z_{i}=0$, $i=0,1,2,3$, we have $\sum_{i}A^{i}_{00}=\sum_{i}A^{i}_{11}=\sum_{i}A^{i}_{22}=\sum_{i}A^{i}_{33}=1$.
For the off-diagonal elements of $A^{i}$,
(i) considering $\sum_{i}\text{e}^{\pm 2\text{i}\varphi_{i}}=0$,
we have $\sum_{i}A^{i}_{03}=\sum_{i}A^{i}_{30}=0$; (ii) considering $\sum_{i}z_{i}\text{e}^{\pm \text{i}\varphi_{i}}=\sum_{i}\text{e}^{\pm \text{i}\varphi_{i}}=0$,
we have $\sum_{i}A^{i}_{01}=\sum_{i}A^{i}_{10}=\sum_{i}A^{i}_{02}=\sum_{i}A^{i}_{20}
=\sum_{i}A^{i}_{13}=\sum_{i}A^{i}_{31}=\sum_{i}A^{i}_{23}=\sum_{i}A^{i}_{32}=0$; and (iii) considering $\sum_{i}z_{i}=0$, we have $\sum_{i}A^{i}_{12}=\sum_{i}A^{i}_{21}=0$.
This means that all of the sum of the off-diagonal elements are exactly zero, i.e., $\sum_{i}A^{i}_{kl}=0, k \neq l$.

By this, we have proved the orthogonality and completeness relations for the present more generalized two-qubit EJM basis $\{|\Phi_{i}\rangle\}, i=0,1,2,3$, parameterized by $z$, $\varphi_{}$ and $\theta_{}$.
Note that here $1/\sqrt{3} \leq |z| \leq 1$ (or $\arcsin (1/\sqrt{3}) \leq \theta_{0} \leq \pi/2$) arises since the restriction $z^{2} = 1/(3\sin^{2}\theta_{0}) \leq 1$.
As an example of our three-parameter basis, if one takes $z=1/\sqrt{3}$ and $\varphi=\pi/4$ then it simplifies to the previous single-parameter EJM \cite{TGB-EJM2021, GuoPRL-EJM2022}.

Consider the reduced states of the present EJM basis states $|\Phi_{i}\rangle$.
By using the result (\ref{reduced-state}), a direct calculation shows that
\begin{eqnarray}
 \langle \Phi_{i}|\vec{\sigma} \otimes I |\Phi_{i}\rangle &=&
\frac{1}{2|z_{}|} \cos \theta
(\sqrt{1-z^{2}}\cos \varphi_{i} + \sqrt{3z^{2}-1}\sin \varphi_{i}, \sqrt{1-z^{2}}\sin \varphi_{i} - \sqrt{3z^{2}-1}\cos \varphi_{i}, z_{i}).
\end{eqnarray}
Let $\varphi_{z}=\text{arg}[(\sqrt{1-z^{2}}+\text{i}\sqrt{3z^{2}-1})/(\sqrt{2}|z|)]$ and thus
\begin{eqnarray}
\cos \varphi_{z} = \frac{1}{\sqrt{2}|z_{}|} \sqrt{1-z^{2}}, ~~
\sin \varphi_{z} = \frac{1}{\sqrt{2}|z_{}|} \sqrt{3z^{2}-1}.
\end{eqnarray}
Then we have
\begin{eqnarray}
\langle \Phi_{i}|\vec{\sigma} \otimes I |\Phi_{i}\rangle =
\frac{1}{\sqrt{2}} \cos \theta
(\cos (\varphi_{i}-\varphi_{z}), \sin (\varphi_{i}-\varphi_{z}), (-1)^{i}/\sqrt{2}), ~~i=0,1,2,3,
\end{eqnarray}
similarly for the reduced states $\langle \Phi_{i}|I \otimes \vec{\sigma}|\Phi_{i}\rangle$.

On the other hand, note that for $z=1/\sqrt{3}$ we have $\vec{m}_{i}=(\sqrt{2}\cos \varphi_{i}, \sqrt{2}\sin \varphi_{i}, (-1)^{i})/\sqrt{3}$, for $z=1/\sqrt{2}$ we have $\vec{m}_{i} =(\cos \varphi_{i}, \sin \varphi_{i}, (-1)^{i})/\sqrt{2}$ and for $z=1$ we have $\vec{m}_{i} =(0, 0, (-1)^{i})$, so these unit vectors $\vec{m}_{i}$s (related to the initial orthogonal states $|\pm m_{i}\rangle, i=0,1,2,3$) can form or not form tetrahedron. By contrast, their reduced states can always form two mirrorimage tetrahedrons inside the Bloch sphere, where the modulus of them are $({\sqrt{3}}/{2})\cos \theta$ and they are pointing toward the unit vectors
$(\sqrt{2}\cos (\varphi_{i}-\varphi_{z}), \sqrt{2}\sin (\varphi_{i}-\varphi_{z}), (-1)^{i})/\sqrt{3}$ or the opposite points.

\subsection{The examples}

We now show several interesting examples by specifying the initial value of parameter $z$ for our three-parameter EJM basis states.

For $z = 1/\sqrt{3}$, we have $\theta_{0} = \arcsin (1/\sqrt{3z^{2}}) = \pi/2$, $z_{0}=z_{2} = 1/\sqrt{3}$, $z_{1}=z_{3} = -1/\sqrt{3}$, $a_{\pm} = \pm 1/\sqrt{3}$, $b_{0,\pm}^{\theta}=b_{2,\pm}^{\theta}=(1 \pm \text{e}^{\text{i}\theta})/\sqrt{6}$ and $b_{1,\pm}^{\theta}=b_{3,\pm}^{\theta}=(-1 \pm \text{e}^{\text{i}\theta})/\sqrt{6}$.
Then we obtain a set of orthonormal basis states
\begin{eqnarray} \label{Phi0-1}
|\Phi_{0, z = 1/\sqrt{3}}\rangle &=& \frac{1}{2}(\text{e}^{-\text{i}\varphi}|00\rangle- r_{+}^{\theta} |01\rangle
                  - r_{-}^{\theta} |10\rangle - \text{e}^{\text{i}\varphi}|11\rangle),
\end{eqnarray}
\begin{eqnarray} \label{Phi1-1}
|\Phi_{1, z = 1/\sqrt{3}}\rangle &=& \frac{1}{2}(-\text{i}\text{e}^{-\text{i}\varphi}|00\rangle + r_{-}^{\theta} |01\rangle
                  + r_{+}^{\theta} |10\rangle - \text{i}\text{e}^{\text{i}\varphi}|11\rangle),
\end{eqnarray}
\begin{eqnarray} \label{Phi2-1}
|\Phi_{2, z = 1/\sqrt{3}}\rangle &=& \frac{1}{2}(-\text{e}^{-\text{i}\varphi}|00\rangle- r_{+}^{\theta} |01\rangle
                  - r_{-}^{\theta} |10\rangle + \text{e}^{\text{i}\varphi}|11\rangle),
\end{eqnarray}
\begin{eqnarray} \label{Phi3-1}
|\Phi_{3, z = 1/\sqrt{3}}\rangle &=& \frac{1}{2}(\text{i}\text{e}^{-\text{i}\varphi}|00\rangle + r_{-}^{\theta} |01\rangle
                  + r_{+}^{\theta} |10\rangle + \text{i}\text{e}^{\text{i}\varphi}|11\rangle),
\end{eqnarray}
described by two real parameters $\varphi \in [-\pi, \pi]$ and $\theta \in [0,\pi/2]$, where $r_{\pm }^{\theta} =(1\pm \text{e}^{\text{i}\theta})/\sqrt{2}$.

For $z = 1/\sqrt{2}$, we have $\theta_{0} = \arcsin (\sqrt{2/3})$, $z_{0}=z_{2} = 1/\sqrt{2}$, $z_{1}=z_{3} = -1/\sqrt{2}$, $a_{\pm} = (\text{i} \pm 1)/2$, $b_{0,\pm}^{\theta}=b_{2,\pm}^{\theta}=(1 \pm \text{e}^{\text{i}\theta})/2$ and $b_{1,\pm}^{\theta}=b_{3,\pm}^{\theta}=(-1 \pm \text{e}^{\text{i}\theta})/2$.
Then we have
\begin{eqnarray} \label{}
|\Phi_{0, z = 1/\sqrt{2}}\rangle &=& \frac{\sqrt{2}-\text{i}}{2\sqrt{3}}
[\frac{(1+\text{i})}{\sqrt{2}}\text{e}^{-\text{i} \varphi}|00\rangle- r_{+}^{\theta} |01\rangle - r_{-}^{\theta} |10\rangle
- \frac{(1-\text{i})}{\sqrt{2}}\text{e}^{\text{i} \varphi}|11\rangle],
\end{eqnarray}
\begin{eqnarray} \label{}
|\Phi_{1, z = 1/\sqrt{2}}\rangle &=& \frac{\sqrt{2}-\text{i}}{2\sqrt{3}}
[\frac{(1-\text{i})}{\sqrt{2}}\text{e}^{-\text{i}\varphi}|00\rangle + r_{-}^{\theta} |01\rangle + r_{+}^{\theta} |10\rangle
- \frac{(1+\text{i})}{\sqrt{2}}\text{e}^{\text{i}\varphi}|11\rangle],
\end{eqnarray}
\begin{eqnarray} \label{}
|\Phi_{2, z = 1/\sqrt{2}}\rangle &=& \frac{\sqrt{2}-\text{i}}{2\sqrt{3}}
[-\frac{(1+\text{i})}{\sqrt{2}}\text{e}^{-\text{i}\varphi}|00\rangle- r_{+}^{\theta} |01\rangle - r_{-}^{\theta} |10\rangle
+ \frac{(1-\text{i})}{\sqrt{2}} \text{e}^{\text{i}\varphi}|11\rangle],
\end{eqnarray}
\begin{eqnarray} \label{}
|\Phi_{3, z = 1/\sqrt{2}}\rangle &=& \frac{\sqrt{2}-\text{i}}{2\sqrt{3}}
[-\frac{(1-\text{i})}{\sqrt{2}}\text{e}^{-\text{i}\varphi}|00\rangle + r_{-}^{\theta} |01\rangle  + r_{+}^{\theta} |10\rangle
+ \frac{(1+\text{i})}{\sqrt{2}}\text{e}^{\text{i}\varphi}|11\rangle].
\end{eqnarray}

For $z = 1$, we have $\theta_{0} = \arcsin (\sqrt{1/3})$, $z_{0}=z_{2} = 1$, $z_{1}=z_{3} = -1$, $a_{\pm} = \text{i}$, $b_{0,\pm}^{\theta}=b_{2,\pm}^{\theta}=(1 \pm \text{e}^{\text{i}\theta})/\sqrt{2}$ and $b_{1,\pm}^{\theta}=b_{3,\pm}^{\theta}=(-1 \pm \text{e}^{\text{i}\theta})/\sqrt{2}$.
Then we have
\begin{eqnarray} \label{}
|\Phi_{0, z = 1}\rangle &=& \frac{1-\sqrt{2}\text{i}}{2\sqrt{3}}
[\text{i}\text{e}^{-\text{i}\varphi}|00\rangle- r_{+}^{\theta} |01\rangle
                  - r_{-}^{\theta} |10\rangle + \text{i}\text{e}^{\text{i}\varphi}|11\rangle],
\end{eqnarray}
\begin{eqnarray} \label{}
|\Phi_{1, z = 1}\rangle &=& \frac{1-\sqrt{2}\text{i}}{2\sqrt{3}}
[\text{e}^{-\text{i}\varphi}|00\rangle + r_{-}^{\theta} |01\rangle
                  + r_{+}^{\theta} |10\rangle - \text{e}^{\text{i}\varphi}|11\rangle],
\end{eqnarray}
\begin{eqnarray} \label{}
|\Phi_{2, z = 1}\rangle &=& \frac{\sqrt{2}\text{i}-1}{2\sqrt{3}}
[\text{i}\text{e}^{-\text{i}\varphi}|00\rangle + r_{+}^{\theta} |01\rangle
                  + r_{-}^{\theta} |10\rangle + \text{i}\text{e}^{\text{i}\varphi}|11\rangle],
\end{eqnarray}
\begin{eqnarray} \label{}
|\Phi_{3, z = 1}\rangle &=& \frac{1-\sqrt{2}\text{i}}{2\sqrt{3}}
[-\text{e}^{-\text{i}\varphi}|00\rangle + r_{-}^{\theta} |01\rangle
                  + r_{+}^{\theta} |10\rangle + \text{e}^{\text{i}\varphi}|11\rangle].
\end{eqnarray}

In fact, by using $\varphi_{z}$ these generalized EJM basis states can also be conveniently expressed as
\begin{eqnarray} \label{Phi0-1-z}
|\Phi_{0}\rangle &=& \frac{1-\text{i}\sqrt{3z^{2}-1}}{2\sqrt{3z^{2}}}
[\text{e}^{-\text{i}(\varphi-\varphi_{z})}|00\rangle- r_{+}^{\theta} |01\rangle
                  - r_{-}^{\theta} |10\rangle - \text{e}^{\text{i}(\varphi-\varphi_{z})}|11\rangle],
\end{eqnarray}
\begin{eqnarray} \label{Phi1-1-z}
|\Phi_{1}\rangle &=& \frac{1-\text{i}\sqrt{3z^{2}-1}}{2\sqrt{3z^{2}}}
[-\text{i}\text{e}^{-\text{i}(\varphi-\varphi_{z})}|00\rangle + r_{-}^{\theta} |01\rangle
                  + r_{+}^{\theta} |10\rangle - \text{i}\text{e}^{\text{i}(\varphi-\varphi_{z})}|11\rangle],
\end{eqnarray}
\begin{eqnarray} \label{Phi2-1-z}
|\Phi_{2}\rangle &=& \frac{1-\text{i}\sqrt{3z^{2}-1}}{2\sqrt{3z^{2}}}
[-\text{e}^{-\text{i}(\varphi-\varphi_{z})}|00\rangle- r_{+}^{\theta} |01\rangle
                  - r_{-}^{\theta} |10\rangle + \text{e}^{\text{i}(\varphi-\varphi_{z})}|11\rangle],
\end{eqnarray}
\begin{eqnarray} \label{Phi3-1-z}
|\Phi_{3}\rangle &=& \frac{1-\text{i}\sqrt{3z^{2}-1}}{2\sqrt{3z^{2}}}
[\text{i}\text{e}^{-\text{i}(\varphi-\varphi_{z})}|00\rangle + r_{-}^{\theta} |01\rangle
                  + r_{+}^{\theta} |10\rangle + \text{i}\text{e}^{\text{i}(\varphi-\varphi_{z})}|11\rangle].
\end{eqnarray}
Here $\varphi_{z=1/\sqrt{3}}=0$, $\varphi_{z=1/\sqrt{2}}=\pi/4$ and $\varphi_{z=1}=\pi/2$ in the above examples. Furthermore, it is not difficult to see that if one takes $\varphi - \varphi_{z} =\pi/4$ then the previous single-parameter EJM basis states \cite{TGB-EJM2021} can be directly obtained, no matter how the parameter $z$ or $\varphi$ is chosen in cylindrical coordinates.
More precisely, given several specified $z$ and $\varphi$, we show the results on the unit vectors $\vec{m}_{i}$s and the reduced states of the corresponding EJM basis states in Table 1.

\begin{table}[tbhp]
\setlength\extrarowheight{0.35\baselineskip}
\begin{threeparttable}
  \caption{Results on the unit vectors $\vec{m}_{i}$s and the reduced states of the corresponding EJM basis states for several specified $z$ and $\varphi$.}
  \label{table1}
  \begin{tabular}{c c c c c c l l}
    \hline
    \hline
    ~$z$~& ~$\varphi_{}$~ & ~$\varphi_{z}$~ &   & ~$z_{i}$~ & ~$\varphi_{i}$~ &~~~~~
    $\vec{m}_{i}$~~~~~~~~~ & ~~ reduced states ~~~\\
      \hline
    \multirow{4}*{$\frac{1}{\sqrt{3}}$}& \multirow{4}*{$\frac{\pi}{4}$} & \multirow{4}*{0}
    & $i=0$ & $\frac{1}{\sqrt{3}}$ & $\frac{\pi}{4}$ & $\frac{1}{\sqrt{3}}(1,1,1)$ & $\pm\frac{1}{2}\cos \theta (1,1,1)$ \\
    \multirow{4}*{}& \multirow{4}*{} & \multirow{4}*{}
    &$i=1$ & $-\frac{1}{\sqrt{3}}$ & $\frac{3\pi}{4}$ & $\frac{1}{\sqrt{3}}(-1,1,-1)$ & $\pm\frac{1}{2}\cos \theta (-1,1,-1)$ \\
    \multirow{4}*{}& \multirow{4}*{} & \multirow{4}*{}
    &$i=2$ & $\frac{1}{\sqrt{3}}$ & $-\frac{3\pi}{4}$ & $\frac{1}{\sqrt{3}}(-1,-1,1)$ & $\pm\frac{1}{2}\cos \theta (-1,-1,1)$ \\
    \multirow{4}*{}& \multirow{4}*{} & \multirow{4}*{}
    &$i=3$ & $-\frac{1}{\sqrt{3}}$ & $-\frac{\pi}{4}$ & $\frac{1}{\sqrt{3}}(1,-1,-1)$ &  $\pm\frac{1}{2}\cos \theta (1,-1,-1)$ \\
    \cline{4-8} 
    \multirow{4}*{$\frac{1}{\sqrt{2}}$}& \multirow{4}*{$\frac{\pi}{2}$} & \multirow{4}*{$\frac{\pi}{4}$}
    & $i=0$ & $\frac{1}{\sqrt{2}}$ & $\frac{\pi}{2}$ & $\frac{1}{\sqrt{2}}(0,1,1)$ &  $\pm\frac{1}{2}\cos \theta (1,1,1)$ \\
    \multirow{4}*{}& \multirow{4}*{} & \multirow{4}*{}
    &$i=1$ & $-\frac{1}{\sqrt{2}}$   & ${\pi}$       & $\frac{1}{\sqrt{3}}(-1,0,-1)$ &  $\pm\frac{1}{2}\cos \theta (-1,1,-1)$ \\
    \multirow{4}*{}& \multirow{4}*{} & \multirow{4}*{}
    &$i=2$ & $\frac{1}{\sqrt{2}}$ & $-\frac{\pi}{2}$ & $\frac{1}{\sqrt{3}}(0,-1,1)$ &  $\pm\frac{1}{2}\cos \theta (-1,-1,1)$ \\
    \multirow{4}*{}& \multirow{4}*{} & \multirow{4}*{}
    &$i=3$ & $-\frac{1}{\sqrt{2}}$   & $0$           & $\frac{1}{\sqrt{3}}(1,0,-1)$ & $\pm\frac{1}{2}\cos \theta (1,-1,-1)$ \\
    \cline{4-8} 
        \multirow{4}*{1}& \multirow{4}*{$\frac{3\pi}{4}$} & \multirow{4}*{$\frac{\pi}{2}$}
    & $i=0$ & 1 & $\frac{3\pi}{4}$ & $(0,0,1)$ & $\pm\frac{1}{2}\cos \theta (1,1,1)$ \\
    \multirow{4}*{}& \multirow{4}*{} & \multirow{4}*{}
    &$i=1$ & $-1$ & $-\frac{3\pi}{4}$ & $(0,0,-1)$ &  $\pm\frac{1}{2}\cos \theta (-1,1,-1)$\\
    \multirow{4}*{}& \multirow{4}*{} & \multirow{4}*{}
    &$i=2$ & 1 & $-\frac{\pi}{4}$ & $(0,0,1)$ &  $\pm\frac{1}{2}\cos \theta (-1,-1,1)$ \\
    \multirow{4}*{}& \multirow{4}*{} & \multirow{4}*{}
    &$i=3$ & $-1$ & $\frac{\pi}{4}$ & $(0,0,-1)$ &  $\pm\frac{1}{2}\cos \theta (1,-1,-1)$ \\
    \hline
    \hline
  \end{tabular}
\end{threeparttable}
\end{table}

\section{Quantum circuits}

We now turn to the quantum circuits for preparing and detecting the present EJM  basis states (\ref{Phi0-1-z})--(\ref{Phi3-1-z}).

\begin{figure}
\centerline{ \Qcircuit @C=1em @R=1em {
&\lstick{\ket{0}} &\gate{H}  &\gate{S} &\ctrl{1} &\gate{X} &\ctrl{1}                           &\qw      &\targ  &\ctrl{1} &\gate{Y} &\ctrl{1}  &\qw \\
&\lstick{\ket{0}} &\gate{H} &\gate{S}  &\gate{R_{y}({\pi}/{2}-2\varphi{'})}   &\qw      &\gate{R^{\dagger}_{}({\pi}/{2}-\theta)} &\gate{H}  &\ctrl{-1}                            &\gate{R^{\dagger}_{}(2\varphi{'})} &\gate{Y} &\gate{S} &\qw \\
}}
\vskip 0.55\baselineskip
\centerline{\footnotesize}
\caption{Quantum circuit for preparing the generalized EJM basis states parameterized by $z$, $\varphi_{}$ and $\theta_{}$, where $\varphi{'}_{}=\varphi_{}-\text{arg}[(\sqrt{1-z^{2}}+\text{i}\sqrt{3z^{2}-1})/(\sqrt{2}|z|)]$.}
\label{circuit-prep.}
\vskip 0.55\baselineskip
\end{figure}
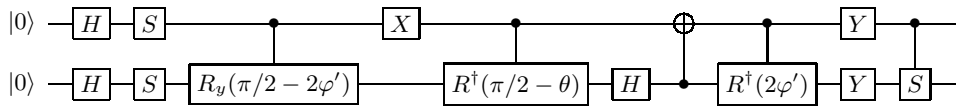

We first consider the quantum circuit for preparing the state (\ref{Phi0-1-z}), as shown in Fig.\ref{circuit-prep.}.
In our quantum circuit, we set the initial two-qubit state $|0\rangle|0\rangle$. It contains several conventionally single qubit gates $H=(\sigma_{x}+\sigma_{z})/\sqrt{2}$, $X=\sigma_{x}$, $Y=\text{i}\sigma_{y}$, $S=\text{diag}(1, \text{i})$ and available two-qubit controlled gates $\text{C}_{R_{y}(\zeta)}$, $\text{C}_{R^{\dagger}_{}(\xi)}$, CNOT gate and controlled-phase gate, where $R_{y}(\zeta)=\text{e}^{-\text{i}\zeta \sigma_{y}/2}$,
$R(\xi)=\text{diag}(1, \text{e}^{\text{i}\xi})$ and $\varphi{'}_{}=\varphi_{}-\varphi_{z}$.
Meanwhile, we note that the basis states $|\Phi_{i}\rangle, i=0,1,2,3,$ can be mutually transformed by local unitary operations
$U_{1}=(I \otimes \sigma_{x})[R(2\varphi{'}+\pi/2) \otimes R^{\dagger}(2\varphi{'}+\pi/2)](\sigma_{x} \otimes I)$ and $U_{2}=\sigma_{z} \otimes \sigma_{z}$,
e.g., $U_{1}|\Phi_{0}\rangle=-|\Phi_{1}\rangle$, $U_{2}|\Phi_{0}\rangle=-|\Phi_{2}\rangle$, and $U_{2}U_{1}|\Phi_{0}\rangle=|\Phi_{3}\rangle$.
So it is possible in principle to prepare any of the basis states characterized by parameters
$z$, $\varphi_{}$ and $\theta_{}$  based on the present quantum circuit by matching unitary operations $U_{1}$ and $U_{2}$.

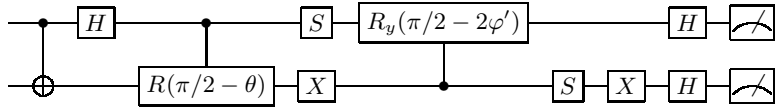
\begin{figure}
\centerline{ \Qcircuit @C=1em @R=1em {
&\ctrl{1} &\gate{H} &\ctrl{1}  &\gate{S}  &\gate{R_{y}({\pi}/{2}-2\varphi{'})} &\qw   &\qw   &\gate{H}   &\meter \\
&\targ    &\qw      &\gate{R_{}({\pi}/{2}-\theta)} &\gate{X}  &\ctrl{-1}                            &\gate{S}  &\gate{X}  &\gate{H}   &\meter \\
}}
\vskip 0.55\baselineskip
\centerline{\footnotesize}
\caption{Quantum circuit for detecting the generalized three-parameter EJM.}
\label{circuit-disc.}
\vskip 0.55\baselineskip
\end{figure}

The quantum circuit for detecting the generalized three-parameter joint measurement is shown in Fig.\ref{circuit-disc.}.
In contrast to the quantum circuit of the previous EJM \cite{TGB-EJM2021}, we here introduce a controlled-rotation operator $R_{y}({\pi}/{2}-2\varphi{'})$ to characterize the present parameters $\varphi_{}$ and $z$.
In this way, the four orthonormal basis states of the present quantum joint measurement can be discriminated in the computational basis. In equations:
\begin{eqnarray} \label{}
|\Phi_{0}\rangle \rightarrow |11\rangle; ~~ |\Phi_{1}\rangle \rightarrow |00\rangle; ~~
|\Phi_{2}\rangle \rightarrow |10\rangle; ~~ |\Phi_{3}\rangle \rightarrow |01\rangle.
\end{eqnarray}

Finally, if one takes $\varphi{'}_{}=\varphi_{}-\varphi_{z}=\pi/4$, then we have $R_{y}({\pi}/{2}-2\varphi{'})=I$ and the present circuit reduces to the previous one \cite{TGB-EJM2021} up to a pair of Pauli-$X$ gates used to remove the global phase.
In particular, if we further take $\theta=\pi/2$ then it naturally simplifies to the standard BSM up to local unitary operations.

\section{Discussion and summary}

In summary, inspired by \cite{Gisin-EJM2019,TGB-EJM2021} we have proposed the more generalized  three-parameter EJM.
We first investigate a set of elegant two-qubit entangled states, characterized by five real parameters $z$, $\varphi_{}$, $\theta_{0}$, $a$ and $\theta_{}$.
There are two main results: (i) the concurrence is only depended on the parameters $a$ and $\theta_{}$, ranging from 0 to 1, and the previous one \cite{TGB-EJM2021} is in the range 1/2 to 1 as the special case for $a=\sqrt{3}$; (ii) two reduced states  are depended on all  parameters and respectively toward the unit vectors $\pm (\sqrt{1-z^{2}}\cos \varphi \sin\theta_{0}+\sin \varphi \cos \theta_{0}, \sqrt{1-z^{2}}\sin \varphi \sin\theta_{0}-\cos \varphi \cos \theta_{0},z\sin \theta_{0})$ shrunk  by $2a\cos \theta /({a^{2}+1})$, having the expected elegant symmetry.
Then we define the more generalized EJM, parameterized by $z$, $\varphi_{}$ and $\theta_{}$, and provide the quantum circuits for preparing and detecting the present EJM basis states by using several available single-qubit gates and two-qubit controlled gates.
In this architecture, we set
$\varphi_{0}=\varphi_{}$, $\varphi_{1}=\varphi_{}+\pi/2$, $\varphi_{2}=\varphi_{}-\pi$, $\varphi_{3}=\varphi_{}-\pi/2$, $z_{0}=z_{2}=z$ and $z_{1}=z_{3}=-z$,
with randomly choosing $\varphi \in [-\pi, \pi]$ and $1/\sqrt{3} \leq |z|\leq 1$.
Interestingly, by contrast, we find that (i) the previous single-parameter EJM \cite{TGB-EJM2021} can be directly obtained by setting $\varphi - \varphi_{z} =\pi/4$, no matter how the initial unit vectors were chosen in cylindrical coordinates, i.e., including
but not limited to the original choice; (ii) not all unit vectors $\vec{m}_{i}$s are suitable for forming EJM, e.g., the parameter $z$ is restricted to $|z| \geq 1/\sqrt{3}$ in our architecture; and (iii) the reduced states of the EJM basis states can always form two mirrorimage regular tetrahedrons inside the Bloch sphere, even without forming a tetrahedron for the initial unit vectors $\vec{m}_{i}$s, robustly preserving its elegant properties.

We have figured out what kind of states the EJM basis belongs to and proposed the more generalized three-parameter two-qubit EJM, which may contribute to studying multi-qubit EJM. However, several open questions hold in our architecture: (i) it is not known whether there exists the more optimized structure so that all the unit vectors in cylindrical coordinates could be used to construct the EJM; (ii) how can such multi-parameter measurement settings be actively applied in the field of quantum information processing; and (iii) how can one generalize the present results to more than two qubits with elegant symmetry?
Finally, we expect that this work will motivate further investigating on the novel EJM and consider its new practical applications, including but not limited to network nonlocalities \cite{TGB-EJM2021,BGT2021IBM,Tavakoli-network2022}, entanglement swapping \cite{GuoPRL-EJM2022} and quantum teleportation \cite{QT-EJM2024}.

\begin{acknowledgements}
This work was supported by
the National Natural Science Foundation of China under Grant Nos: 62271189, 12071110,
the Hebei Central Guidance on Local Science and Technology Development Foundation of China under Grant Nos: 226Z0901G, 246Z0902G,
the Hebei 3-3-3 Fostering Talents Foundation of China under Grant No: A202101002,
the Education Department of Hebei Province Natural Science Foundation of China under Grant No: ZD2021407,
the Education Department of Hebei Province Teaching Research Foundation of China under Grant No: 2021GJJG482.
\end{acknowledgements}

\end{document}